\documentclass[epj]{svjour}
\usepackage{graphicx}
\usepackage{here}
\begin{document}
\title{Near threshold production of the pseudoscalar mesons at the COSY-11 facility}
\author{P.~Moskal\inst{1,2}, R.~Czy{\.z}ykiewicz\inst{1,2}, H.-H.~Adam\inst{3}, 
        A.~Budzanowski\inst{4},
        E.~Czerwi\'nski\inst{1}, D.~Gil\inst{1}, D.~Grzonka\inst{2}, 
        M.~Janusz\inst{1,2}, L.~Jarczyk\inst{1}, B.~Kamys\inst{1}, 
        A.~Khoukaz\inst{3},
        K.~Kilian\inst{2}, P.~Klaja\inst{1,2}, W.~Oelert\inst{2},
        C.~Piskor-Ignatowicz\inst{1}, J.~Przerwa\inst{1,2}, B.~Rejdych\inst{1},
        J.~Ritman\inst{2}, T.~Sefzick\inst{2}, 
        M.~Siemaszko\inst{5}, M.~Silarski\inst{1}, J.~Smyrski\inst{1},
        A.~T\"aschner\inst{3}, M.~Wolke\inst{2}, P.~Winter\inst{6},
        P.~W\"ustner\inst{2}, M.~J.~Zieli\'nski\inst{1}, W.~Zipper\inst{5}
}                     
\institute{Institute of Physics, Jagellonian University, 30-059 Cracow, Poland \and
           IKP \& ZEL, Forschungszentrum J\"ulich, 52425 J\"ulich, Germany \and
           IKP, Westf\"alische Wilhelms-Universit\"at, 48149 M\"unster, Germany \and
           Institute of Nuclear Physics, 31-342 Cracow, Poland \and
           Institute of Physics, University of Silesia, Katowice, Poland \and
           Department of Physics, University of Illinois at Urbana-Champaign, Urbana, IL 61801, USA
}
\date{Received: date / Revised version: date}
%
\abstract{We summarise  measurements of the COSY-11 collaboration
concerning the excitation functions of the near threshold pseudoscalar meson production
in the proton-proton interaction. The results are discussed in the context of the meson-nucleon
and hyperon-nucleon interactions. We conclude that the $\eta$-proton interaction is significantly stronger 
than the $\eta^{\prime}$-proton interaction. Similarly, we found that the hyperon $\Lambda$ interacts with the nucleon
considerably stronger than the hyperon $\Sigma$,  and that the interaction of $K^-$-proton 
is much stronger than this of the K$^+$-proton.
} 
\authorrunning{P.~Moskal et al.}
\titlerunning{Near threshold production of the pseudoscalar mesons at the COSY-11 facility}
\maketitle
\section{Introduction}
In the low energy regime of the quantum chromodynamics,
where the interaction 
between quarks and gluons cannot be treated 
perturbatively, 
hadrons become the 
relevant degrees of freedom and the knowledge 
of their interactions is of the basic importance. 
In this article we will present results of the investigations
of the meson-nucleon and hyperon-nucleon interactions
based on the 
shape of the near threshold total cross section 
excitation functions  for the production of the hyperons and pseudoscalar mesons.
Due to the space limitation we will restrict the discussion to the conclusions
derived from the total cross sections of the proton-proton collisions only.

\section{K$^+$K$^-$ and K-proton interaction}
An important feature of near-threshold measurements 
is connected with the fact that due to the proximity of bound or quasi-bound states 
of some of the reaction products, interaction between them can be very strong, thus 
influencing the measured cross sections essentially. This creates an opportunity 
to investigate interaction between particles which cannot be accessed in 
direct elastic scattering experiments. For example, measurements of 
the reaction $pp \to pp K^{+}K^{-}$ allow one to investigate the 
kaon-antikaon interaction. Such measurements can help us to understand 
the nature of the scalars  $f_{0}(980)$ and $a_{0}(980)$ which 
masses are very close to the mass of a kaon pair. The nature of these 
objects remains a long-standing problem of the meson physics. The standard 
quark model has difficulties with interpreting these mesons as 
quark-antiquark pair. Therefore, various 
non $q-\bar{q}$ descriptions have been proposed 
including a four-quark system~\cite{jaffe}, 
or a kaon-antikaon molecule~\cite{hanhart,weinstein,lohse}. 
Especially for the formation of 
the molecule, the strength of the kaon-antikaon interaction 
is of the crucial importance.
For study of this interaction we have carried out measurements of the 
$pp \to ppK^{+}K^{-}$ reaction near the kinematical threshold. 
The measurements 
were conducted  
using the 
COSY-11 apparatus~\cite{brauksiepe,jurek,klaja} 
and the cooler synchrotron COSY~\cite{prasuhn}. 
They were based 
on kinematically complete reconstruction of positively charged ejectiles 
while the negative kaon was identified 
via the missing mass~\cite{wolke,quentmeier,winter}. 
\begin{figure}[h]
  \includegraphics[width=.45\textwidth]{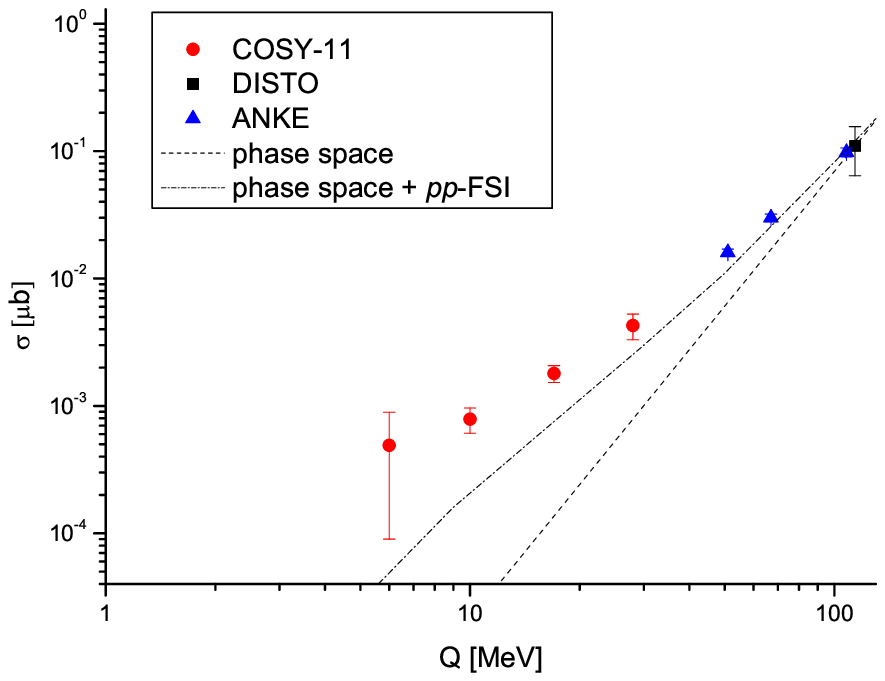}
  \vspace{-0.3cm}
 \caption{ Total cross section for the reaction
            $pp \to pp K^+K^-$~\cite{wolke,quentmeier,winter,maeda,kesh,balestra} as a function of the excess energy Q.
            The COSY-11 data points~\cite{wolke,quentmeier,winter} 
            lie significantly above the expectations indicated by the lines that
            are normalized to the data point measured by the DISTO collaboration~\cite{balestra}.
            \label{crosskk}}
\end{figure}
Our results combined  with the data obtained by 
DISTO~\cite{balestra} and ANKE~\cite{maeda,kesh} collaborations
are clearly showing that towards the lower values of the excess energy Q 
the total cross section value is exceeding expectations 
based on the homogeneous phase space population including 
the $pp$ final state interaction (FSI), 
as it can be seen in Figure~\ref{crosskk}. The observed difference might originate from the 
$pK$ and/or $KK$ FSI. We investigated the effect of the interaction 
between particles in the final state using also distributions of invariant 
masses of $pK^{+}$, $pK^{-}$ and $K^{+}K^{-}$ pairs~\cite{winter}. 
Within the limited 
statistics the distribution for the $pK^{-}$ pairs shows an enhancement 
towards lower masses. Moreover, the enhancement remains when the distribution
of the $pK^{-}$ invariant mass is normalized to the corresponding  $pK^{+}$ spectrum,
indicating that the $pK^{-}$ interacts  much stronger than the $pK^{+}$~\cite{winter,colinc11}.
Indeed, in the recent article of the ANKE collaboration it is shown 
that most of  the close-to-threshold enhancement of the excitation
function can be explained by the $K^-$-proton interaction. 
However, the data points closest to the threshold are still
above this predictions indicating the signal 
from the KK or $K^+$-proton interaction.
For a strict description of the final state, 
calculations based on application of the four-body formalism are 
required. 
%
%
%
%
%
%
%
%
\section{Hyperon-nucleon interaction}
The existence of light hypernuclei, such as $^{3}$He$_{\Lambda}$, 
shows the low energy $\Lambda$-p interaction to be 
strongly attractive, though not sufficient to bind 
the hyperon-deuteron~\cite{balewskiepj}. The hyperon-nucleon 
interaction is of special interest since it is  
influenced by the strange quark content of the 
hyperon. However, in contrast to the nucleon-
nucleon case, due to the short lifetime of  
hyperons, the direct measurements of low-energy 
hyperon-nucleon scattering are sparse and the 
resulting parameters are rather poorly known~\cite{balewskiepj}.\\
Using the COSY-11 facility
we have determined the 
excitation function of the $pp \to p K^{+}\Lambda$,
$pp \to p K^{+}\Sigma^{0}$ 
and $pp \to nK^{+}\Sigma^{+}$ reactions in the near threshold 
energy range. 
Surprisingly, the total cross section for the 
production of the hyperon $\Lambda$ was found to be by a 
factor of thirty larger than this for $\Sigma^{0}$. It is in 
drastic contrast to the results of the cross section 
ratio $\sigma(pp\to pK^{+}\Lambda)/\sigma(pp\to pK^{+}\Sigma^{0})$ determined at 
higher energies, where it was found to be equal to 
three as expected from the isospin relations. This 
observation raised an interesting question 
whether the drastic increase of the cross section 
ratio near threshold is a mere effect of the $\Lambda$-p 
interaction or whether it is due to the reaction 
mechanism. To explain this unexpected increase 
 different models have been proposed based on 
the coherent exchange of the $\pi$ and K mesons~\cite{haiden} or 
on the excitation of the intermediating 
resonances~\cite{sib1,sib2,shyam1,shyam2}. 
All these models failed however to predict the 
value of the total cross section for the $pp \to n K^{+}\Sigma^{+}$ 
reaction. To understand the hyperon-nucleon 
interaction a further thorough theoretical 
investigations are needed. 
\begin{figure}[ht]
  \includegraphics[width=0.45\textwidth]{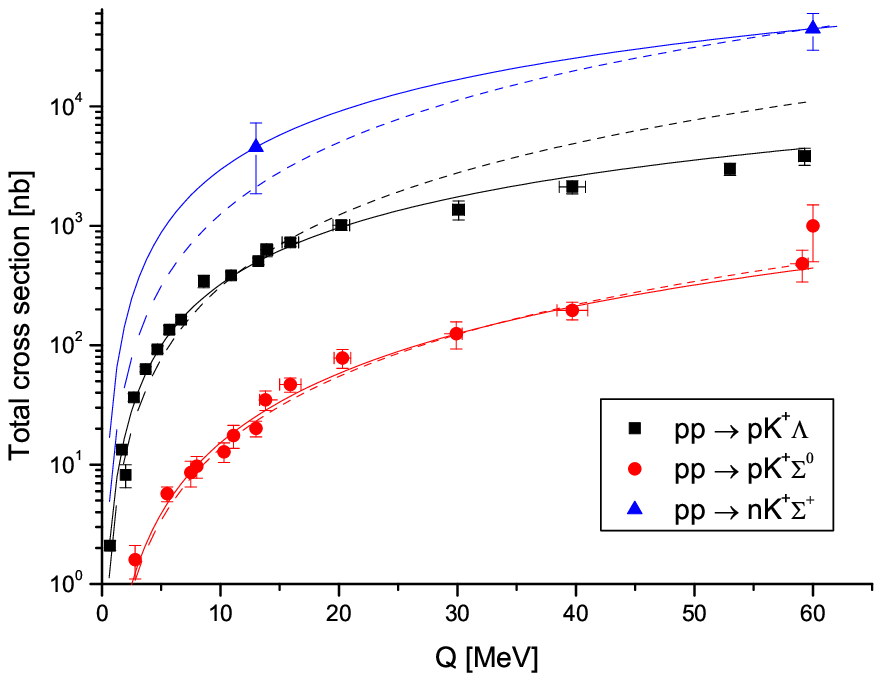}
  \vspace{-0.3cm}
 \caption{Total cross section as function of the  excess energy Q for the near threshold production
           of the hyperons $\Lambda$, $\Sigma^0$ and $\Sigma^+$ via the
           $pp\to pK^+\Lambda$~\cite{balewski,sewerin,kowina,bilger,abd}, $pp\to pK^+\Sigma^0$~\cite{sewerin,kowina,bilger,fritsch},
           and $pp\to nK^+\Sigma^+$~\cite{rozek} reactions, respectively.
           \label{crosshyperons}}
\end{figure}
Figure~\ref{crosshyperons} presents the COSY-11 data together with 
expectations derived under the assumption of the 
homogeneously populated phase space~(dashed lines) and the 
phase-space modified by the hyperon-nucleon 
interactions~(solid lines). The comparison of the calculations 
with the data suggests much weaker final-state 
interaction in the p-$\Sigma^{0}$ channel than in the case of 
the p-$\Lambda$. 
Therefore, the most plausible seems to be an explanation that the observed enhancement
of the production of the $\Lambda$ over the $\Sigma$ hyperon
is predominantly due to the large $\Lambda$-proton interaction and relatively 
negligible $\Sigma$-proton interaction. 
This would also explain why the $\Lambda$-hypernuclei are observed and 
there are no $\Sigma$-hypernuclei~\cite{colinc11}

Regarding the
\mbox{n-$\Sigma^{+}$}  interaction due to the 
large 
systematic uncertainties, and only two available data points,
any conclusions would be at present premature.
\section{Interaction of the $\eta$ and $\eta^{\prime}$ mesons with protons}
In 
this section we give account of the studies of the 
interactions between the $\eta$ and $\eta$' mesons with 
nucleons. It is rather challenging to conduct such 
research because these mesons decay within a 
distance of tens of femtometers rendering their 
direct detection impossible. It is also completely 
unfeasible  to accomplish out of them a beam or a 
target. Therefore, we have produced these mesons 
in the collisions of protons close to the kinematical 
threshold  where the outgoing particles possess 
low relative velocities and remain in the distance 
of few femtometers long enough to experience the 
strong interactions which may manifests itself in a 
measurable manner.\\
\vspace{-1.0cm}
\begin{figure}[H]
  {\centerline{\includegraphics[width=.47\textwidth]{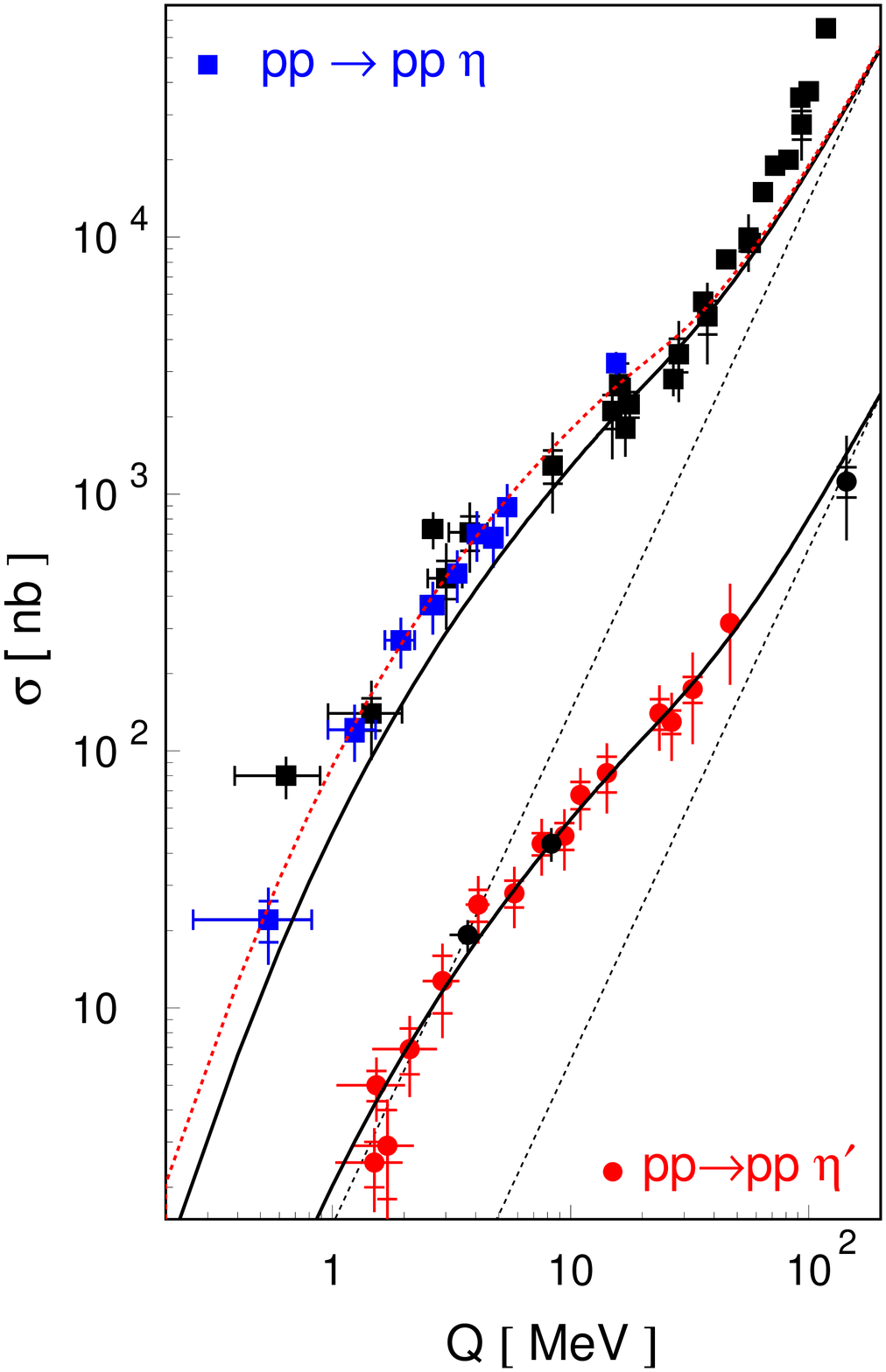}}}
 \caption{ Total cross section as a function of the excess energy Q for the reactions
    $pp\to pp\eta$~\cite{hibou,chiavassa,calen1,bergdolt,calen2,smyrski,moskalprc} (squares) and
    $pp\to pp\eta^{\prime}$~\cite{balestra2,wurzinger,moskal1,moskal2,khoukaz} (circles). The results determined using
    the COSY-11 facility and the synchrotron COSY (red and blue) are shown together with the data
    from the CELSIUS and SATURNE facilities (black). The dashed lines indicate a phase space integral
    normalized arbitrarily. The solid lines show the homogeneous phase space ditribution with inclusion
    of the proton-proton strong and Coulomb interaction. Results of calculations taking into account
    additionally the interaction between the $\eta$ meson and the proton
    is presented by the red dotted line.
    \label{crosseta}}
\end{figure}
Measurements of the 
$pp\to pp \eta$ and $pp\to pp \eta^{\prime}$ reactions have been conducted
close to the 
kinematical threshold. The obtained near threshold excitation functions are presented in Figure~\ref{crosseta}.
The remarkable difference 
between the shape of the excitation function of 
the $pp\to pp \eta$ and $pp\to pp \eta^{\prime}$ reactions allowed to 
conclude that the interaction between the $\eta^{\prime}$ 
meson and the proton is significantly weaker than 
the analogous $\eta$-proton interaction~\cite{hab,swave}. 
This is the 
first ever empirical appraisal of this hitherto 
entirely unknown force. 
Using the COSY-11 facility we have also determined 
two-particle invariant mass  distributions 
for the production of both the $\eta$ and the $\eta^{\prime}$ meson.
The  data were taken at the same excess energy in order to enable 
the direct comparison of the experimental distributions~\cite{hab}.
The final results will be reported elsewhere in the near future.
\section{Acknowledgment}
We acknowledge the support of the
European Community-Research Infrastructure Activity
under the FP6 programme (Hadron Physics, 
RII3-CT-2004-506078), the support
of the Polish Ministry of Science and Higher Education under the grants
No. 
3240/H03/2006/31  and 1202/DFG/2007/03,
and  the support of the German Research Foundation (DFG).

\end{document}